\documentclass[aps, twocolumn,noshowpacs,preprintnumbers,amsmath,amssymb]{revtex4}

\usepackage{verbatim}
\usepackage{wasysym}

\usepackage{graphicx}
\usepackage{dcolumn}
\usepackage{bm}
\usepackage{wrapfig}

\begin{document}

\title{Field-effect transistors made of individual colloidal PbS nanosheets}

\author{Sedat Dogan}
\author{Thomas Bielewicz}
\author{Yuxue Cai}
\author{Christian Klinke}
\affiliation{Institute of Physical Chemistry, University of Hamburg, Hamburg, Germany}

\begin{abstract} 

Two-dimensional materials are considered for future quantum devices and are usually produced by extensive methods like molecular beam epitaxy. We report on the fabrication of field-effect transistors using individual ultra-thin lead sulfide nanostructures with lateral dimensions in the micrometer range and a height of a few nanometers as conductive channel produced by a comparatively fast, inexpensive, and scalable colloidal chemistry approach. Contacted with gold electrodes, such devices exhibit p-type behavior and temperature-dependent photoconductivity. Trap states play a crucial role in the conduction mechanism. The performance of the transistors is among the ones of the best devices based on colloidal nanostructures.

\end{abstract}

\maketitle

Inexpensive electronic applications require semiconductor materials which can be easily processed, e.g. by spin-coating or dip-coating \cite{1}.  Thus, researchers are looking for materials that are solution processable while exhibiting reasonable electronic properties. Colloidal semiconductor nanoparticles are among the candidates to be integrated into low-cost electronic devices \cite{2}. They are suspended in liquid media, mass-producible, and tunable in their optical and electrical properties due to quantum confinement effects \cite{3}.  Colloidal nanomaterials are promising due to the simplicity and thus the inexpensiveness of their production and subsequent processing. One hurtle which needs to be overcome is the presence of tunnel barriers in the nanoparticle films which lead to high resistances. This effect is the consequence of long isolating organic ligands capping the nanoparticles´ surface. These ligands can be either replaced by shorter ones including halides \cite{4},  or removed by physicochemical processes \cite{5}.  These post-treatments deteriorate the nanoparticle surface but reduce the resistive power losses. A different approach to reduce the tunnel barriers consists in the use of inorganic capping "ligands" such as In$_{2}$Se$^{2-}$ on CdSe nanoparticles \cite{6}.  Such films find applications e.g. as field-effect transistors \cite{7},  thermoelectrics \cite{8},  and photoconductors \cite{9}. 

Yet another approach is to avoid tunnel barrier from the beginning on and to synthesize continuous two-dimensional materials in solution. Indeed, some progress has been made in controlling the lateral dimensions \cite{10}  and thickness \cite{11} of nanostructures through varying parameters like the nature of the ligands which are used to bind to specific facets of the nanocrystals and inhibit an isotropic growth \cite{12}.  Recently, we demonstrated that two-dimensional PbS nanosheets can be produced in solution by colloidal chemistry \cite{13}.  We showed that lead sulfide nanosheets form due to two-dimensional oriented attachment of small zero-dimensional colloidal nanocrystals. The nanosheets have a height of a few nanometers and exhibit lateral dimensions in the order of a micron. Nevertheless, the control of anisotropic growth in nanocrystal syntheses is still a great challenge. The PbS nanosheets used in the here presented study were synthesized based on the recently reported procedure with small changes \cite{10}. One modification made compared to the reported synthesis was that tri-n-octylphosphine (TOP) was omitted from the lead-precursor. 860 mg of lead acetate was dissolved in a mixture of 10 mL diphenylether (DPE) and 3.5 mL oleic acid (OA). The blend was heated under N$_{2}$ atmosphere to 85$^{\circ}$C and degassed for about 2 hours. Then, the mixture was heated to 110$^{\circ}$C and 1 mL 1,1,2-trichloroethane (TCE) was added rapidly before finally at 130$^{\circ}$C a suspension of 12 mg thioacetamide (TAA) in 930 $\mu$L TOP and 70 $\mu$L dimethylformamide (DMF) was added to start the reaction. After 5 minutes the mixture was slowly cooled down to room temperature and centrifuged. The residue was washed twice with toluene and the final nanosheet material was re-suspended in toluene. Figure 1a shows a transmission electron microscopy image of a PbS nanosheet synthesized in this way. By the omittance of TOP in the complexation of the lead precursor the nanosheets become more stable and grow larger in lateral dimensions compared to the ones previously reported. This makes the handling and further measurements of the physical properties of the nanosheets easier. 

\begin{figure}[ht]
  \centering
  \includegraphics[width=0.4\textwidth]{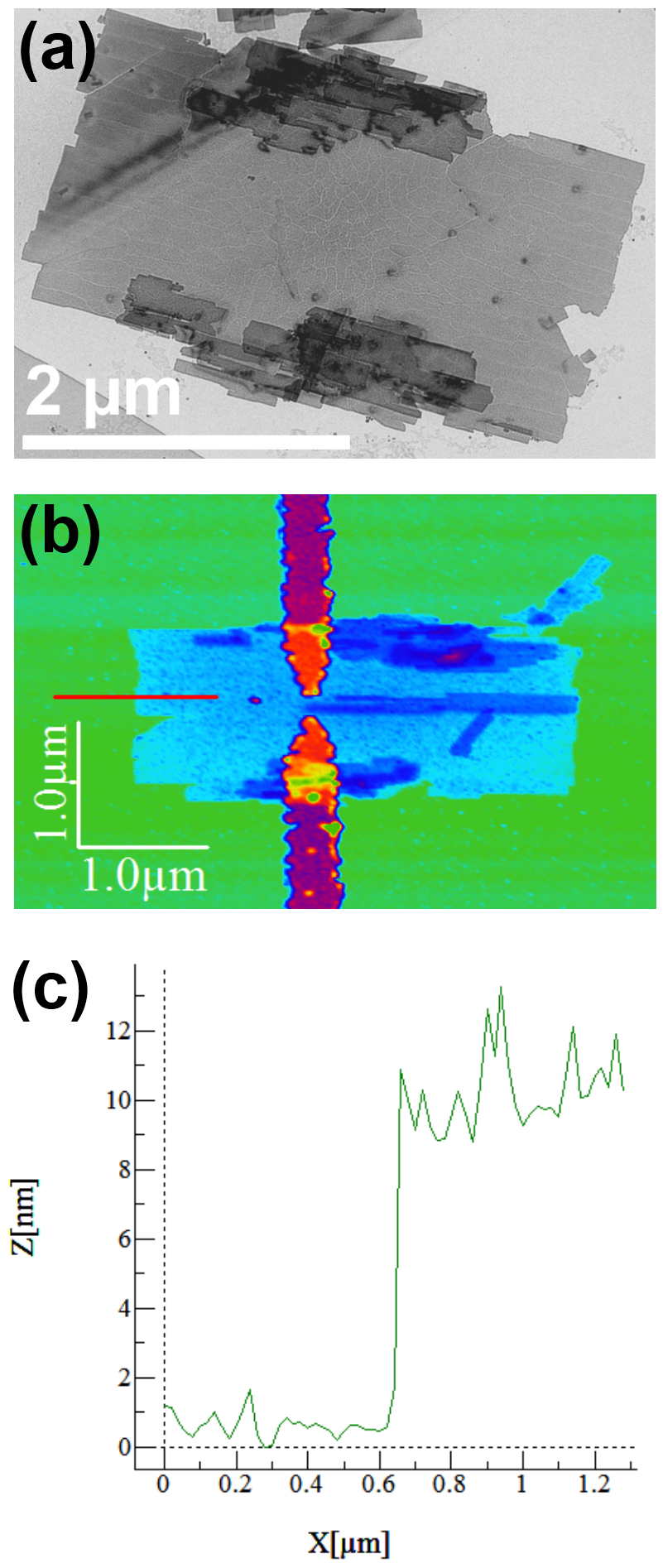}
  \caption{\textit{(a) TEM image of an ultra-thin single layer PbS nanosheet, (b) AFM image of an individual nanosheet contacted with two gold electrodes, (c) Height profile across the nanosheet marked as red line in (b). Considering that the nanosheet is covered on both sides by a monolayer of oleic acid with a length of 1.8 nm, the thickness of the inorganic nanosheet is about 5.4 nm.}}
\end{figure}

In order to characterize their electrical properties the toluene suspended PbS nanosheets were spread by spin-coating on a silicon substrate with 300 nm thermal oxide. The silicon is highly n-doped and is used as back gate. Subsequently, gold electrodes (1 nm Ti + 40 nm Au) were defined by electron-beam lithography on individual nanosheets resulting in a field-effect transistor configuration. Figure 1b shows an atomic force microscopy image of two electrical contacts on a PbS nanosheet taken after the electrical characterization. The nanosheet has a lateral size of 1.5 $\mu$m x 3.5 $\mu$m, and a total height of about 9 nm including a top and a bottom layer of self-assembled oleic acid of a thickness of about 1.8 nm, yielding a height of the inorganic part of about 5.4 nm (Figure 1c). 

Immediately after contacting, the samples were transferred to a probe station connected to a semiconductor parameter analyzer. The measurements were performed under vacuum conditions. In Figure 2a the output characteristics of the device (shown in Figure 1b) are displayed for bias voltages VDS between -2.5 V and +2.5 V (Vg = 0 V) and at temperatures between 77 K and room temperature. The inset shows the same data plotted on a logarithmic scale. The curves unveil that at room temperature at any voltage different from 0 V a current is flowing and the device has a finite zero-bias conductance, which indicates that a good contact of the gold electrodes to the PbS nanosheet is established. The asymmetry regarding positive and negative bias values is ascribed to the asymmetry in the contacts as can be seen in Figure 1b. At lower temperatures the curves show again a nonlinear behavior with a distinct asymmetry. The zero-bias conductivity drops to the noise level. This is an indication for barriers at the contacts which cannot be thermally overcome at lower temperatures. The non-linearity of the curves is probably also due to barriers at the contacts. At lower temperatures these circumstances lead to a substantially suppressed current at lateral voltages below 1 V. 

\begin{figure}[ht]
  \centering
  \includegraphics[width=0.4\textwidth]{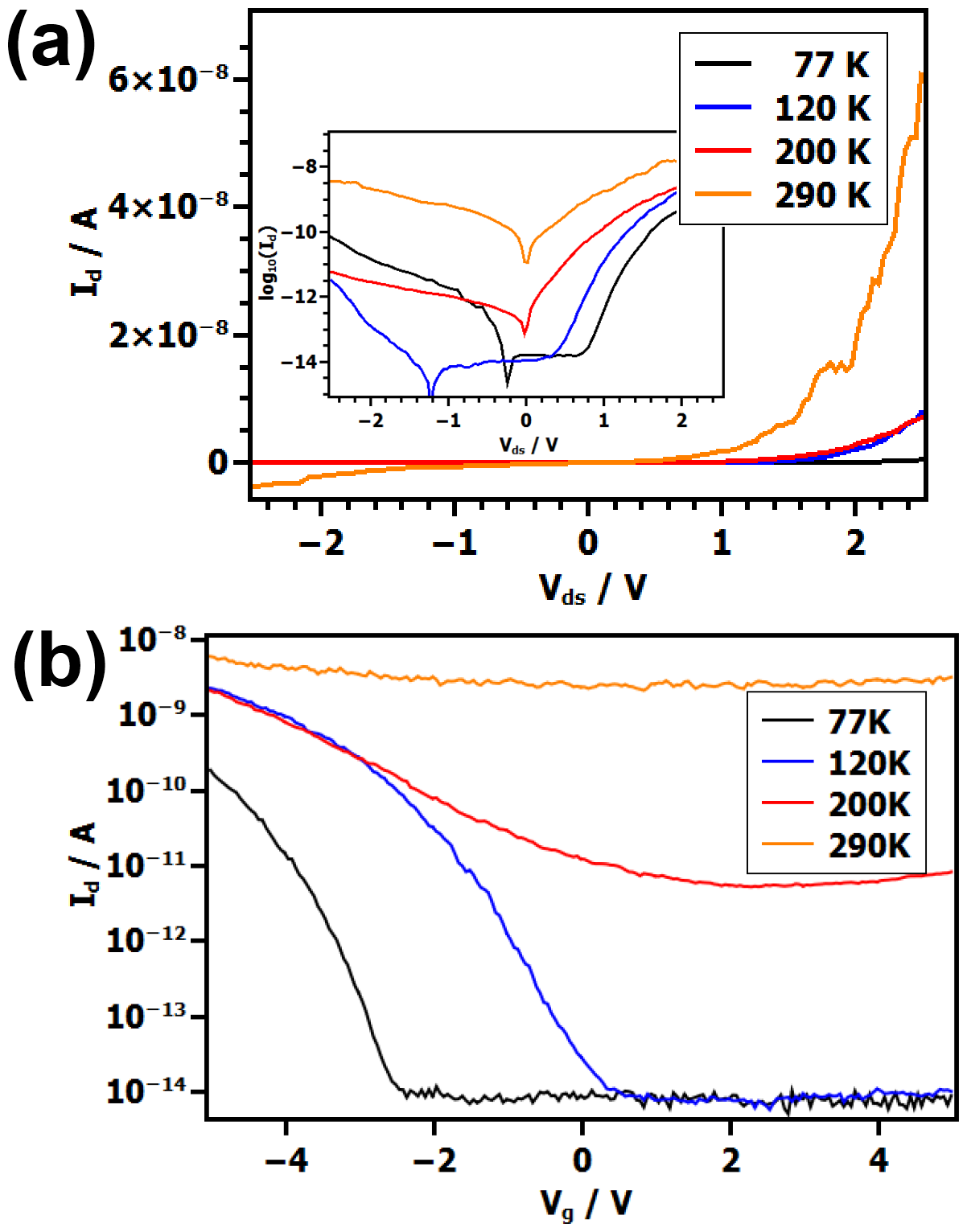}
  \caption{\textit{(a) Output characteristics at room and lower temperatures for a fixed gate voltage of Vg = 0 V. Inset: Same data as log(IDS)-VDS. (b) Transfer characteristic at a source-drain voltage of VDS = +1 V.}}
\end{figure}

The devices were also characterized by sweeping the gate voltage from -5 V to +5 V (VDS = +1 V) yielding the transfer characteristics shown in Figure 2b. The current is higher for negative gate voltages, which indicates that holes are the majority carriers in the conductive channel of the device. This means that the devices show p-type behavior, though it is not very pronounced at room temperature and at this range of gate voltages. It can be seen that although the current is less for lower temperatures, the curves are smoother and the switching is more pronounced. It becomes possible to completely switch off the devices to the noise level of the measurement setup. The ON/OFF ratio amounts to 3.93·105 at 120 K with a sub-threshold slope of 532 mV/dec, whereas at room temperature the ON/OFF ratio is only 2.82. 

Another way to evaluate the transistor performance is by the figure-of-merit mobility. The field-effect mobility $\mu$, which indicates how fast charge carriers move inside the material upon application of an electrical field, can be determined by the expression

\begin{equation}
\mu_{FE} = \frac{dI_{DS}}{dV_{g}} \cdot \frac{L^{2}}{V_{DS} C}
\end{equation}

where $L$ is the device channel length between source and drain electrodes and $C$ the capacitance between the channel and the back gate. The channel length $L$ is 160 nm and the channel width $W$ is 100 nm. The semiconductor channel is separated from the back gate by a SiO$_{2}$ layer of $d$ = 300 nm. Thus, the capacitance is calculated by 

\begin{equation}
C = \epsilon_{0} \epsilon_{r} \cdot \frac{L \cdot W}{d}
\end{equation}

with $\epsilon_{r}$(SiO$_{2}$) = 3.9. Then, the maximum field-effect mobility of the device shown here calculates to remarkable 0.417 cm$^{2}$V$^{-1}$s$^{-1}$ (Vg = -5 V, VDS = 1 V, T = 290 K). Furthermore, the conductivity has been calculated by 

\begin{equation}
\sigma = \frac{I_{DS}}{V_{DS}} \cdot \frac{L}{h \cdot W}
\end{equation}

to 72.06 mS/cm (VDS = 2.5 V, Vg = 0 V, T = 290 K), with the height of the nanosheet $h$ = 5.4 nm. These values are minimum values for this material, since no measures had been taken to improve these values, e.g. by chemical or thermal treatments. The mobility and the conductivity of the pristine material are already larger than state-of-the-art thin film field-effect transistors using PbSe \cite{14}  or PbS \cite{15}  nanocrystals. Our approach does not require post-treatments, and the continuity of the two-dimensional nanosheets assures the seamless charge transport through the whole channel material. We measured 36 devices with similar output and transfer characteristics.

\begin{figure}[ht]
  \centering
  \includegraphics[width=0.4\textwidth]{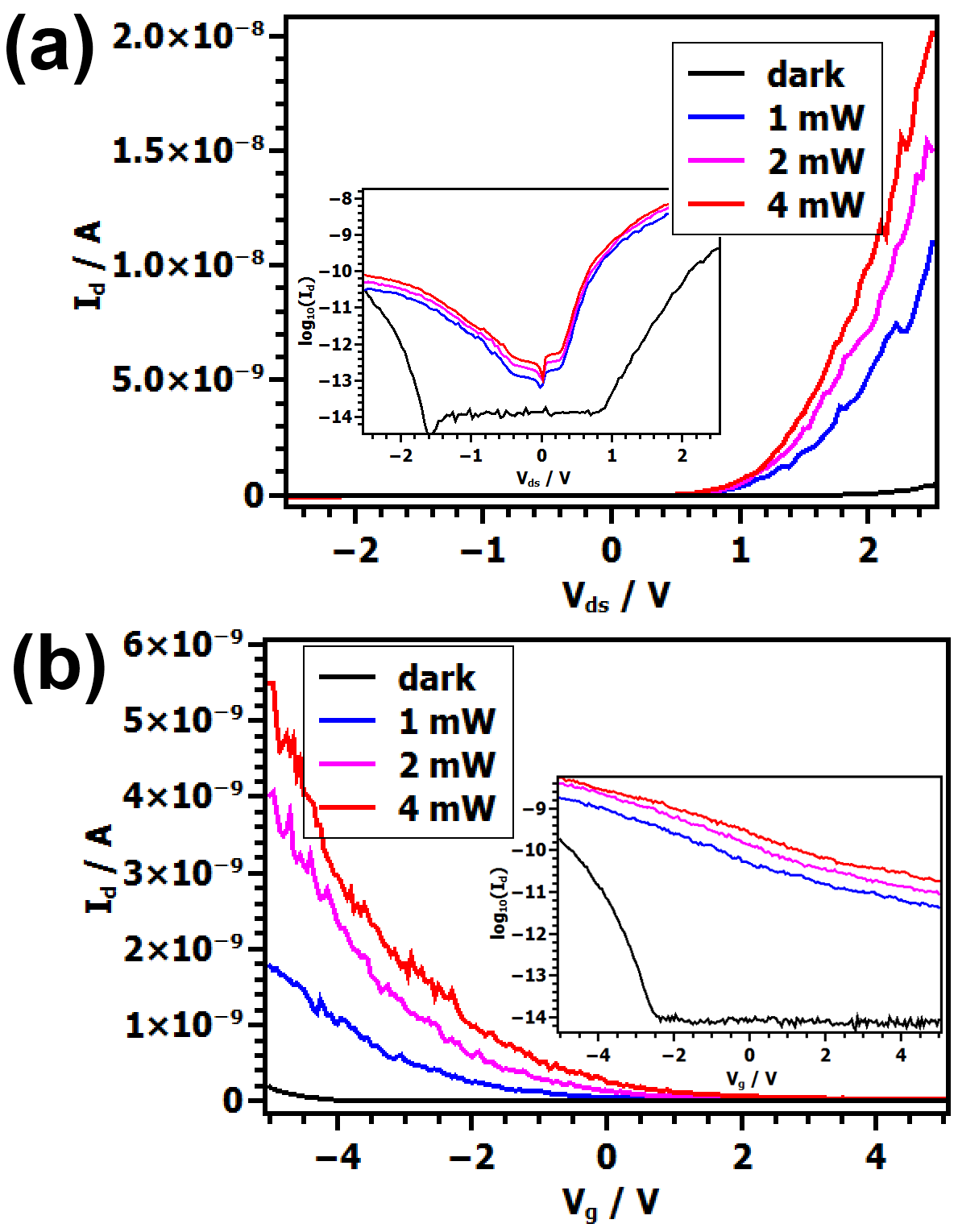}
  \caption{\textit{Output (a) and transfer (b) characteristics in darkness and under laser illumination (637 nm) with different intensities at 77 K.}}
\end{figure}

Beside the surprising fact that two-dimensional materials can be formed in cubic systems, PbS nanosheets show also interesting opto-electric properties \cite{13}. As a semiconductor with a direct bandgap in the infrared region, PbS is a good photo-absorber material \cite{16}.  From bulk semiconductor systems to low-dimensional semiconductors the physical properties change significantly, which is due to quantum confinement of the charge carriers. In two-dimensional semiconductor systems the average thickness is smaller than the Fermi wavelength in z direction, the height \cite{17}.  Thus, the transistor characterization has been performed at different temperatures with and without illumination using a 637 nm laser. At room temperature the effect of illumination was moderate. However, upon illumination of the devices at lower temperatures, as shown for 77 K in Figure 3, a pronounced increase in current was observed. At 77 K the current increased by a factor of 7404 for an intensity of 1 mW/cm$^{2}$ (VDS = 1 V, Vg = 0 V), whereas under the same conditions at 290 K only a factor of 1.8 was reached. An increase in intensity increased also the current flowing. Interestingly, the strongest increase was observed in the region below 1 V where the conductivity in darkness is virtually zero. This can also be seen in the transfer characteristic in Figure 3b.

\begin{figure}[ht]
  \centering
  \includegraphics[width=0.4\textwidth]{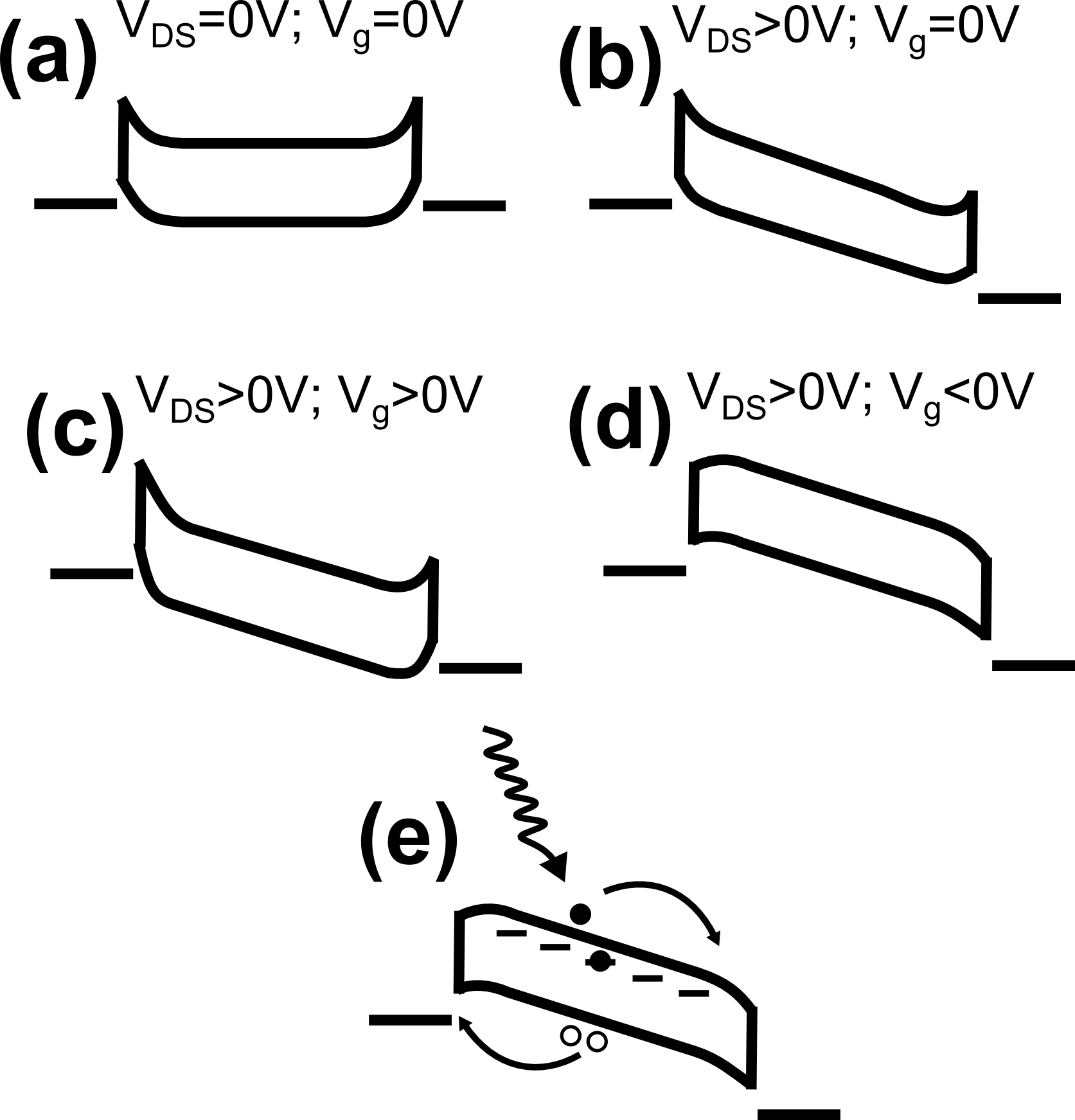}
  \caption{\textit{Conduction mechanism.}}
\end{figure}

The bandgap of bulk PbS is 0.41 eV. By the simple "particle-in-a-box" approach the effective bandgap in a confined PbS nanosheet of a thickness of 5.4 nm can be calculated to be 0.63 eV. Since the effective masses of electron and hole are almost equal it can be assumed that the conduction and valence band shift by the same amount. Considering an electron affinity of about 4.1 eV for confined PbS makes it possible to draw its energy levels \cite{18}.  In spite of the confinement the valence band of the PbS nanosheets is still above the work function of the gold electrodes of 5.1 eV. Thus, after Fermi level equilibration Schottky barriers will be formed at the Au-PbS contacts for the electrons and Ohmic contacts for the holes (Figure 4a). This will promote a current based on holes as majority carriers (Figure 4b). Positive values for the gate voltage reduce the hole current and favor the electron current (Figure 4c), whereas negative gate voltages increase the hole current (Figure 4d). 

The illumination of the samples lifts additional electrons to the conduction band which fill trap states and photo-generated carrier harvesting becomes possible. This is especially very efficient at around zero bias or zero gate voltage. At lower temperatures this is more effective since the trap states will not depopulate and phonon scattering is reduced (Figure 4e). For negative gate voltages the absolute value of the photo-current increases more than for positive gate voltages, but the relative increase is higher for positive gate voltages. The origin of this behavior can be explained by the electrical field associated band bending. When the gate voltage is negative the resulting band bending at the contacts is efficient for hole transport. Additionally, photo-generated holes are efficiently transported to the source contact. On the other side, at the drain contact the electrons are collected. Since in the dark the electron transport is suppressed the photo-generated electrons contribute efficiently to the increase in current. The generation of additional charge carriers leads partially to an increase in current due to charge carrier harvesting, but it mainly leads to filling of trap states which in turn increases the conductivity of the material. 

In summary, we demonstrated that gold contacted pristine PbS nanosheets work as transistors with p-type character and are suited for photo-electric applications. Especially at lower temperatures the relative increase of photo-current is remarkable. The pristine nanosheets could be used as easy-processable, inexpensive semiconductor material exhibiting mobilities and conductivities outperforming the best state-of-the-art colloidal materials, without any additional chemical or thermal treatment.


\end{document}